\magnification=1200 \baselineskip=13pt
\hsize=16.5 true cm \vsize=20.5 true cm
\def\parG{\vskip 10pt} \def\parGG{\vskip 15pt}
\font\bbold=cmbx10 scaled\magstep2
\font\slarg=cmti10 scaled\magstep1
\font\blarg=cmbx10 scaled\magstep1
\font\larg=cmr10 scaled\magstep1

\centerline{{\slarg Physics Education} {\blarg 35},
{\larg 110 (March 2000)}}\parGG

\centerline {\bbold Pin-Hole Water Flow from Cylindrical
Bottles}\parGG
Paulo Murilo Castro de Oliveira, Antonio Delfino,\par
Eden Vieira Costa and Carlos Alberto Faria Leite\parG
Instituto de F\'\i sica, Universidade Federal Fluminense\par
av. Litor\^anea s/n, Boa Viagem, Niter\'oi RJ, Brazil 24210-340\par
e-mail PMCO @ IF.UFF.BR\par

\vskip 1cm\leftskip=1cm\rightskip=1cm 

{\bf Abstract} We performed an experiment on elementary
hydrodynamics. The basic system is a cylindrical bottle from which
water flows through a pin-hole located at the bottom of its lateral
surface. We measured the speed of the water leaving the pin-hole, as
a function of both the time and the current level of water still
inside the bottle. The experimental results are compared with the
theory. The theoretical treatment is a very simple one based on mass
and energy conservation, corresponding to a widespread exercise
usually adopted in university basic disciplines of Physics.\par

We extended the previous experiment to another similar system using
two equal bottles with equal pin-holes. The water flowing from the
first bottle feeds the second one located below it. The same concepts
of mass and energy conservation now lead to a non-trivial
differential equation for the lowest bottle dynamics. We solved this
equation both numerically and analytically, comparing the results
with the experimental data.\parGG

\leftskip=0pt\rightskip=0pt\vfill\eject

\centerline{\bf I --- Introduction}\parGG

Many university textbooks (see, for instance, [1--7]) refer to the
problem of water squirted freely from a small hole in a bottle, as in
figure 1. The water speed $V$ at the hole depends on the height $H$
of the free liquid surface above the pin-hole, according to the
so-called Torricelli's law

$$V^2 = 2 g H\,\,\,\, ,
\eqno(1)$$

\noindent where $g$ represents the earth gravitational constant. This
result was obtained by Galileo's assistant E. Torricelli, in 1636 [7].
It is normally obtained nowadays through Bernoulli's theorem

$$p + {1\over 2} \rho V^2 + \rho g y = C\,\,\,\, ,
\eqno(2)$$

\noindent introduced by Daniel Bernoulli in his book on hydrodynamics
of 1738, a century later than Torricelli's law. Here, $\rho$
represents the liquid density, while $p$, $V$ and $y$ are
respectively the local hydrostatic pressure, speed and height, all of
them measured at the same point inside the liquid. The constant $C$
means that the sum on the left-hand side is the same, independent of
the particular point where the measurements are performed. In our
case, one must take two points: the first at the free liquid surface
inside the bottle, with $y = H$, neglecting the speed there; and the
second point just after the pin-hole, with $y = 0$, outside the
bottle. The hydrostatic pressure at these two points is the same,
namely the atmospheric one.  Equation (1) follows directly. This
theorem holds for stationary liquid flows, and comes from energy
conservation arguments. Thus, in adopting it one is neglecting any
energy loss due to viscosity, turbulence, etc.

Because the downward speed of the free liquid surface inside the
bottle was neglected, there is a missing factor of $(1 - a^2/A^2)$ on
the left-hand side of equation (1), where $A$ and $a$ represent the
cross sections of the bottle and the liquid vein, respectively. This
correction comes from equation (7) to be introduced later on. The
cylindrical bottle used in our experiment has a diameter of 9.5cm,
and the pin-hole is made of a thin plastic tube (extracted from a
ball pen) whose diameter is 2.0mm. Thus, the neglected factor of $(1
- a^2/A^2)$ represents a relative deviation of the order of
$10^{-7}$, much smaller than our experimental accuracy. Also, the
liquid vein cross section $a$ is a little bit smaller than that of
the pin-hole, due to the phenomenon of {\it vena contracta} [3,5,7].

The text is organized as follows. In section II we describe our
first, simplest experiment with a single bottle (figure 1), where we
measured simultaneously the speed $V$ and the time $T$ spent since
the initial height $H_0$, while the bottle drains off, for successive
values of the decreasing height $H$. The speeds are indirectly
obtained through the range $X$ measured on the horizontal rule
located at a fixed height below the bottle: $X$ is thus proportional
to $V$. In order to compare our experimental results with the theory,
a simple correction is made on Torricelli's law, taking energy losses
into account. We also obtain the solution of Bernoulli's differential
equation (2) for this case, giving the time evolution of $V(T)$ and
$H(T)$ during the flow, comparing them with the experiment. In
section III we describe our second experiment with two identical
bottles (figure 2), the water squirted from the first one being
caught by the other through a funnel. Initially, the bottles are
equally filled, and then both flows start at the same time. Since we
have already verified the validity of Torricelli's relation with good
accuracy from the first experiment, we used it in the second one,
instead of measuring the bottom velocity $V'$. Our accuracy in
measuring $H$ (or $H'$) is better than the corresponding one for $X$
(or $X'$, or $V'$). Thus, in this second experiment we measured only
the time evolution of $H'(T)$. In this case, the analytical solution
for Bernoulli's differential equation is not trivial, and we solved
it numerically.  Nevertheless, the analytical solution could be
obtained by using some tricks. The results are compared with the
experiment.  In section IV we present our conclusions.

\vskip 30pt
\centerline{\bf II --- One-Bottle Experiment}\parGG

The water is squirted through a horizontal thin plastic tube
installed at the pin-hole. A (parallel) rule is placed below it at a
distance $Y_0$. Thus, the falling time of a small piece of water is
$\sqrt{2Y_0/g}$. This is the constant of proportionality between the
measured range $X$ in figure 1 and the speed $V$, i.e. $V =
X/\sqrt{2Y_0/g}$. The water flow is started by one of the authors,
responsible for measuring heights, from an initial value $H_0 =
20.0$cm, at time $T = 0$. At this same instant, another author
starts a chronometer with memories (actually, a computer program
which stores the time spent since $T = 0$ each time some key is
stroked), and a third author reads $X$ at the rule. A fourth author
books this value. The whole procedure is repeated when the liquid
level reaches the mark of $H = 19.5$cm, again at $H = 19.0$cm, then
at $H = 18.5$cm, and so on. At the end, one has a three-column table
with $H$, $T$ and $X$ (this last being proportional to the speed
$V$). The initial range $X_0 = (28.0 \pm 0.3)\,$cm was measured for
$Y_0 \cong 11$cm (for the sake of clarity, figure 1 shows a larger
value, but our actual measures were all taken with $Y_0 \cong 11$cm),
correponding to an initial speed $V_0 \cong 1.9$m/s, supposing $g
\cong 9.8$m/s$^2$. However, one does not need to perform the
transformation from $X$ to $V$: we will use always $X$ instead of
$V$. In doing so, we do not care about precise measures of $Y_0$ and
$g$.

Figure 3 shows the plot of the squared range $X^2$ {\it versus} the
height $H$, which must be a straight line, namely $X^2 = 4 Y_0 H$,
according to Torricelli's law, equation (1). However, our
experimental straight line does not cross the origin! On the
contrary, the best linear fit for our data, the continuous line $X^2
= K (H-R)$ with optimum values for $K$ and $R$, crosses the $H$ axis
at a minimum residual height $R = 2.05$cm. Indeed, during the
experiment, we noted that the continuous water flow ceases {\sl
before} the surface level inside the bottle reaches the pin-hole.
This precociously interrupted flow occurs when the height is still
around 2cm. At this situation, the surface tension inside the thin
tube is enough to compensate alone the overpressure due to the
residual level height, except for isolated drops which starts to
appear after this point. Even before this, for heights below 5cm, we
could note that the water flow leaving the pin-hole is no longer
completely stable as it was for the initial heights above 5cm: some
oscillations and instabilities appear below that height, as commented
later on.

Anyway, equation (1) is not supposed to fit the experimental data,
for which we cannot neglect energy losses. Indeed, the sum $C$ in
equation (2) is not strictly the same for all positions inside the
liquid, but must be smaller at the pin-hole exit than inside the
bottle. This behavior is expected due to the viscosity and
turbulences leading to energy losses when the liquid passes through
the thin tube. When applying Bernoulli's equation (2) for a point at
the free liquid surface and another point at the pin-hole exit, one
must discount some energy density amount from the former one. We will
assume here that this amount is the same during all the flow, which
corresponds to subtract a constant from the kinetic energy term in
equation (2), or, alternatively, subtract a constant height $R$ from
the actual height $H$. With this correction, Torricelli's law becomes

$$V^2 = 2 g (H-R)\,\,\,\, ,\eqno(3)$$

\noindent instead of equation (1), where the residual height $R$ is a
constant. Indeed, our experimental data agree quite well with this
modified form of Torricelli's law, equation (3), as can be
appreciated in figure 3. Also from figure 3, one can measure the
slope $K = 4Y_0 = 43.36$cm, in complete agreement with the rule's
vertical distance $Y_0 = 10.8$cm.

The subtracted term $R = 2.05$cm corresponds to an energy density
$\rho g R$ dissipated at the thin tube, i.e. the lost energy per unit
volume of liquid. Thus, considering the outflow rate $a V$, i.e. the
volume of liquid crossing the tube per unit time, one obtains an
energy dissipation rate, i.e. a dissipated power $\rho g a R V$
proportional to the speed $V$. This energy dissipation comes from the
fact that different points correspond to different liquid speeds:
small pieces of water near the internal walls, inside the thin tube,
move slower than those near the tube axis. The relative displacement
between adjacent liquid layers dissipates the energy (cylindrical
layers inside the thin tube). The constancy of $R$ during the whole
flow, experimentally supported by the straight line behavior in
figure 3, means that the energy dissipation occurs according to a
constant resultant viscous force $\rho g a R \approx 6 \times
10^{-4}$N. Of course, the particular value of this force must depend
on the liquid's viscosity, and also on the characteristics of the
thin tube itself, e.g. its diameter and length, the material of which
it is made, etc. Thus, the value of this constant must change for
another experimental apparatus. In order to test this, we performed
the same experiment once again, by using another bottle, now with an
inner diameter of 5.1cm and initial height $H_0 = 42$cm, also with
another pin-hole. The experimental results are shown in figure 4. The
straight line behavior agrees again with our modified Torricelli's
law, equation (3). The thin tube is now metalic, with an inner
diameter of 3.5mm. Accordingly, the new value $R = 4.81$cm
corresponds to a larger resulting viscous force of $5 \times
10^{-3}$N, approximately 7 times larger than that of our first
bottle. However, this new value is again constant during the whole
flow. For this new bottle, we start to observe oscillations and
instabilities of the outflow below $H \approx 12$cm. For $H \approx
8$cm these instabilities become strong enough to forbid accurate
measurements of $X$.

It is possible to make our results independent of the particular
geometry of the (cylindrical) bottle used during the experiment,
provided the condition $a << A$ remains valid. In order to do such an
universal (bottle-independent) analysis, instead of the quantities
$H$ and $V$ denoted by capital letters, we will adopt the following
lowercase, dimensionless, reduced variables

$$h = {H-R \over H_0-R}\,\,\,\, ,\eqno(4)$$

\noindent and

$$v = {V\over V_0} = {X\over X_0}\,\,\,\, .\eqno(5)$$

\noindent With these definitions, the modified Torricelli's law,
equation(3), reads simply

$$v = \sqrt{h}\,\,\,\, .\eqno(6)$$

\noindent Figure 5 shows a plot of the experimental values of $v$
{\it versus} $h$, as well as the continuous line corresponding to the
theoretical equation (6). Data from both bottles, presented first
separately in figures 3 and 4, are now superimposed into the same
plot, defining a single, universal, bottle-independent curve. Except
for small deviations at the end of the flow, near the origin in
figure 5, the agreement between theory and experiment is quite good.
A better test for Torricelli's law is shown in figure 6, where the
same experimental data were plotted in a $log \times log$ scale. The
continuous line is the best linear fit for our first bottle
(circles), excluding data taken for $H < 5$cm, where the quoted flow
instabilities and oscillations start to be noted by us during the
experiment. The slope of 0.497 we obtained must be compared with the
exponent $1/2$ of the square root appearing in equation (6). For the
other bottle (crosses), excluding $H < 12$cm, we obtained a slope of
0.501.

Hereafter, all our remainder experimental results correspond to the
bottle geometry shown in figure 1, because the measures are more
accurate in this case (note the number of experimental points in
figure 3, more than twice the corresponding number in figure 4). When
the liquid vein crosses the rule height $Y_0$ below the pin-hole, it
is not completely stable, presenting some fluctuations as visible in
figure 1. In measuring $X$, we are forced to estimate an ``average''
value at each measuring time. Our actual measurements were made at a
vertical height $Y_0 \cong 11$cm, smaller than that shown in figure
1, in order to minimize these fluctuations. Indeed, the flow seems
very stable near the pin-hole, at least for large values of the
height $H$. Once we have already a good agreement between theory and
experiment, concerning the modified Torricelli's law, hereafter we
will use equations (4) and (6) in order to obtain the reduced
velocity $v$ from the actually measured height $H$, instead of
equation (5) with the measured horizontal range $X$. The reason for
that is the better accuracy we have in measuring $H$ at the marks
previously glued on the bottle surface, compared with $X$ read at the
horizontal rule. From now on, we will study the complete flow
dynamics instead of only the instantaneous Torricelli's relation
between speed and height.

Figure 7 exhibits the reduced speed $v$ as a function of the time $T$
elapsed since the initial height $H_0 = 20.0$cm. The continuous line
is the best linear fit, excluding the last six points where
instabilities at the pin-hole start to be noted. According to this
fit, the complete flow spends a time $T_0 = 526$s. The real time,
however, is smaller than that, something around 450s, according to
the last point in figure 7. For $H < 5$cm the flow at the pin-hole
starts to present small intermittent bursts superimposed to the
continuous, stable flow observed before. This phenomenon is more and
more clearly observed as time goes by, until only isolated drops
could be observed at the end, below the residual height $R$, with no
longer traces of any underlying continuous flow. This
continuous-plus-intermittent way is more effective in draining off
the bottle faster than it would be by following the previously
observed pure-continuous flow, according to the straight line in
figure 7. Also, Bernoulli's theorem, even with our constant energy
amount subtracted at the pin-hole, is no longer valid: it supposes
one has a {\sl stationary} flow. The theoretical treatment of this
intermittency problem is a very difficult task, and many interesting
complex phenomena could be observed experimentally [8]. No
satisfactory analytical approach is available, although some computer
simulations based on a stochastic simple model seems to capture the
essential physical ingredients governing the phenomenon [9] --- see
[10] for a review.

Before the intermittent regime arises below $H \approx 5$cm, however,
Bernoulli's theorem can be applied to our bottle flow, in order to
compare the results with the dynamical experimental counterparts
shown in figure 7. One also needs to include the conditon that the
liquid is not compressible, i.e.

$$ a V = - A {dH\over dT}\,\,\,\, ,\eqno(7)$$

\noindent where $a$ and $A$ are the cross sections already mentioned.
The result is a simple differential equation, namely

$${dV\over dT} = - {a\over A} g\,\,\,\, .\eqno(8)$$

\noindent According to this result, the speed $V$ (or $v$) decreases
linearly as time goes by, just as can be seen in the experimental
plot, figure 7, before the intermittent regime starts to appear. Also
according to this result, the complete draining time would be [5]

$$T_0 = {AV_0\over ag} = {A\over a} \sqrt{2(H_0-R)\over g}\,\,\,\, .
\eqno(9)$$

\noindent Taking our experimental values $T_0 = 526$s, $H_0 = 20.0$cm
and $R = 2.05$cm, adopting $g \cong 9.8$m/s$^2$, and considering a
bottle diameter of 9.5cm, we can estimate $a$. As already mentioned,
$a$ is not the cross section of the pin-hole, but that of the liquid
vein ({\it vena contracta}). Indeed, equation (9) gives a diameter of
1.8mm for the {\it vena contracta}, a little bit smaller than that of
the pin-hole, namely 2.0mm.

From now on, we will also use a lowercase, dimensionless, reduced variable

$$t = {T\over T_0}\,\,\,\, ,\eqno(10)$$

\noindent when referring to the time. With this notation, the
theoretical time evolutions of the reduced speed and height are
simply

$$v = 1 - t\,\,\,\, ,\eqno(11)$$

\noindent and

$$h = (1 - t)^2\,\,\,\, ,\eqno(12)$$

\noindent again independent of the particular (cylindrical) bottle
actually used during the experiment. Both the experimental and
theoretical values for these evolutions are shown in figure 8. Once
more, deviations between theory and experiment are visible at the
end, where the flow starts to present an intermittent, non-stationary
component. Note, in particular, that the final height ($R \cong 2$cm
within the particular geometry of our bottle, corresponding to $h =
0$ in figure 8) is actually reached faster than it would be according
to the completely stable flow observed before (theoretical,
continuous curve). The intermittency appearing at the end of the
whole process helps to drain off the bottle faster.

\vskip 30pt
\centerline{\bf III --- Two-Bottles Experiment}\parGG

Considering now the system shown in figure 2, in applying condition
(7) to the lowest bottle, one needs to take into account the extra
amount of water coming from the upper bottle. It reads now

$$ a (V' - V) = - A {dH'\over dT}\,\,\,\, ,\eqno(13)$$

\noindent where the height $H'$ and the speed $V'$ refer to the
lowest bottle, being related one to the other by the same modified
Torricelli's law

$${V'}^2 = 2 g (H'-R)\,\,\,\, ,\eqno(14)$$

\noindent already verified in our first experiment, with $R =
2.05$cm. Both the bottles and the pin-hole tubes are identical. We
define the reduced speed

$$u = {V'\over V_0}\,\,\,\, ,\eqno(15)$$

\noindent analogously to equation (5). The time evolution of $u$
follows the dimensionless differential equation

$$(1 + {{\rm d}u\over {\rm d}t})\, u = 1 - t\,\,\,\, ,\eqno(16)$$

\noindent which is the same equation (13), after being simplified
by using (14), (15), (9) and (11). Here, $t$ is the same reduced
time already defined in equation (10).

One can solve this equation numerically, by dividing the time
interval $0 \leq t \leq 1$ in $N$ equal subintervals. The lowercase
indices $n = 0, 1, 2 \dots N$ represent the discretized times at the
borders of these subintervals, i.e. $t_n = n/N$, while $u_n$
represents the value of $u$ at time $t_n$. The derivative in equation
(16) will be replaced by the approximation

$${{\rm d}u\over {\rm d}t} \approx {u_{n+1} - u_n\over 1/N}\,\,\,\,
,\eqno(17)$$

\noindent where $1/N$ is the size of each subinterval. This
approximation is as good as we want, provided a large value for $N$
is adopted. In other words, the numerical solution we will obtain is
exact in the sense that one always can keep the errors below any
predefined tolerance, no matter how exacting is the user who
chooses the degree of tolerance. Equation (17) stands for the
derivative of $u$ taken at the center of the subinterval, i.e. at
time

$$t = {n+1/2\over N}\,\,\,\, ,\eqno(18)$$

\noindent which is the value replacing $t$ at the right-hand side of
equation (16). Accordingly, another approximation, namely

$$u \approx {u_{n+1} + u_n\over 2}\,\,\,\, ,\eqno(19)$$

\noindent will replace $u$ outside the parenthesis at the left-hand
side of (16). Then, by performing these replacements, equation (16)
is transformed into a second degree equation for $u_{n+1}$ whose
solution is

$$u_{n+1} \approx {1\over 2N} \big[ \sqrt{(2Nu_n-1)^2 + 8(N-n) - 4} - 1
\big]\,\,\,\, .\eqno(20)$$

\noindent Now, one needs only to choose a convenient value for $N$,
program this equation on a computer (or even a pocket calculator) and
process it starting from $u_0 = 1$ with $n = 0$ at the right-hand
side. The result will be the next value, i.e. $u_1$. Then, processing
again the same program with this recently obtained value for $u_1$
and $n = 1$ at the right-hand side, one gets $u_2$. After repeating
this process $N$ times, one has the complete function $u(t)$ along
the interval $0 \leq t \leq 1$.

The lowest bottle is fed by the upper one, up to $t = 1$ when the
upper flow ceases. However, at this time the lowest bottle is still
flowing with a reduced speed $u^*$. From this moment on, there is
only one flowing bottle, as in our first experiment. Thus, $u$ will
decrease linearly, starting with $u = u^*$ at $t = 1$, at the same
constant rate observed in our first experiment, namely ${\rm d}u/{\rm
d}t = -1$. The bottom flow finally also ends at $t = 1 + u^*$. The
value $u^*$ is then a key parameter for our system. By running our
program for equation (20) with $N = 100$, we obtained $u^* =
0.54630191$. Running it again with $N = 1,000$, we got $u^* =
0.54629310$, which means that $N = 100$ is enough for an accuracy up
to the fourth decimal figure, corresponding to a relative deviation
of the order of $10^{-5}$, far below our experimental accuracy. By
running the same program with $N = 10,000$, and once again with $N =
100,000$, which spends much less than one second on a PC, we obtained

$$u^* = 0.54629302\,\,\,\, ,\eqno(21)$$

\noindent in both cases, meaning that this value is the exact one, at
least up to the $8^{\rm th}$ decimal figure.

From equation (16) one can verify two limiting cases. First, near $t
= 0$, one must have $u \cong 1 - t^2/2$. Note that the first
derivative of $u$ vanishes at $t = 0$, as it must be at the beginning
because the outflux of the lowest bottle is exactly compensated by
the influx coming from the top bottle (remember that both bottles
start from the same initial heights $H_0 = H'_0$). Second, in
reaching $t = 1$ from below, one has ${\rm d}u/{\rm d}t = -1$,
compatible with the one-bottle constant flow regime which will be
installed from this moment on. Both limiting cases can also be
observed in equation (20), namely $u_1 \cong 1 - 1/2N^2$ and $u_N -
u_{N-1} \cong -1/N$.

Alternatively, one can solve equation (16) analytically, by replacing
$u$ by $\omega(1-t)$, which allows one to separate the variables $t$
and $\omega$ into different sides of the resulting equation, i.e.

$${{\rm d}t\over 1 - t} = {\omega {\rm d}\omega\over 1 - \omega +
\omega^2}\,\,\,\, .\eqno(22)$$

\noindent Performing the integration, one finds finally the ugly
analytical solution

$$1 - t = {\exp{ \big( -{\pi\over 6\sqrt{3}} \big)}
\exp{ \big[ -{1\over\sqrt{3}} {\rm tg}^{-1}\big({2\omega-1\over
\sqrt{3}}\big) \big ]} \over \sqrt{\omega^2 - \omega + 1}}\,\,\,\,
.\eqno(23)$$

\noindent Going back to $u = \omega(1 - t)$, we have plotted $u$ {\it
versus} $t$ from this solution, a task which requires much more
computer time than the whole processing of the numerical solution
(20). The resulting plot is indistinguishable from the one obtained
numerically with $N = 100$. Also, by taking the limit $t \to 1$ in
this solution, we obtained

$$u^* = \exp\big( -{\pi\over 3\sqrt{3}} \big)\,\,\,\, ,\eqno(24)$$

\noindent in complete agreement with the numerically obtained value
(21).

The above paragraphs of this section concern what says the theory
about our two-bottles experiment. We have also measured the
successive heights $H'$, from which we obtained the reduced speed

$$u = \sqrt{H' - R\over H_0 - R}\,\,\,\, ,\eqno(25)$$

\noindent as a function of the time. The result is plotted in figure
9, together with the theoretical continuous curve. The point denoted
by P corresponds to $u^*$, at $t = 1$. Our experimental value is

$$u^* = 0.535\,\,\,\, ,\eqno(26)$$

\noindent obtained by interpolating the two experimental points which
are nearest to $t = 1$. The relative deviation between theory and
experiment, concerning this value, is only $2\%$. In performing the
experiment, we have been careful enough to interrupt the upper flow
as soon as the top bottle reaches the residual level $R \cong 2cm$.
After this, only isolated drops would feed the lowest bottle, but
they were not included in our one-bottle, continuous flow analysis
which ends at the residual level $R$. Thus, this further flow from
isolated drops cannot be included now into our two-bottles
experiment.

Note again deviations at the end of the flow, similar to those
already appeared in the one-bottle experiment: the last points are
located below the theoretical straight line, because the intermittent
flow regime accelerates the draining off. Note also the smaller
deviations appearing just before $t = 1$, now the experimental points
being {\sl above} the theoretical curve. This is so because the
accelerated draining off of the top bottle, due to the intermittent
flow occurring just before $t = 1$, feeds water into the lowest
bottle {\sl faster} than it would be according to the theoretical,
continuous flow.

\vskip 30pt
\centerline{\bf IV --- Conclusions}\parGG

We performed a simple experiment on elementary hydrodynamics,
verifying the validity of the so-called Torricelli's law, equation
(1), relating the height $H$ of the free surface of water in a bottle
with the speed $V$ at which the liquid is squirted through a
pin-hole. The speeds are indirectly measured through the horizontal
range $X$ (figure 1). In order to take energy losses into account, a
simple correction was introduced into this law: equation (3) replaces
the original form (1). According to our experimental observations
(figures 3 and 4) the residual height $R$ subtracted from $H$ in
order to take energy losses into account is a constant during all the
flow. We have also measured the dynamical evolution of this system,
i.e. the dependences of both $H$ and $V$ on the time $T$, comparing
the results with the theoretical framework of Bernoulli's theorem. We
performed also another experiment, using two equal bottles, the
liquid squirted by the first feeding the second one. In this case, we
got a non-trivial differential equation from the same framework. It
was solved both numerically and analytically, and the results were
compared with the experimental dynamical evolution.

We believe such simple experiments could be reproduced by students,
in order to better understand elementary hydrodynamic concepts. Also,
other similar systems --- e.g. the ones quoted as examples, exercises
and problems in references [5--7] --- could be experimented along the
same lines presented here. In such experiments, the fixed parameters
characterizing each bottle are: 1) the initial height $H_0$; 2) the
initial speed $V_0$; 3) the residual height $R$, experimentally
obtained through the plot $V^2$ {\it versus} $H$ (figures 3 and 4);
and 4) the total time $T_0$ required to drain off the bottle,
experimentally obtained through the plot $V$ {\it versus} $T$ (figure
7). On the other hand, the dynamical variables depending on the time
$T$ during the flow are: $H$ and $V$ for our one-bottle experiment
(figure 1); or $H'$ and $V'$ for our two-bottles experiment (figure
2).  Nevertheless, neither the expressions of these dynamical
variables as functions of the time nor the mathematical relation
between themselves would depend on the particular geometry of the
cylindrical bottles and pin-holes actually used in the experiments.
In order to stress this independence, we adopted the reduced,
dimensionless, bottle independent variables $h$, $v$, $t$ and $u$
defined by equations (4), (5), (10) and (15). According to these
dimensionless variables, the theoretical equations (6), (11), (12),
(20) and (23) can be experimentally tested with other bottles,
different from those we used here. We also believe the students must
be introduced as often as possible to the very good scientific
practice of using universal variables, independent of particular
implementations or parameters.

This work was partially supported by brazilian agencies CAPES, CNPq
and FAPERJ. One of us (PMCO) is grateful to the late professor Pierre
Henrie Lucie from whom he learned how to use simple systems and
experiments in order to teach Physics. Other (AD) is grateful to J.S.
S\'a Martins and Nivaldo A. Lemos, for enlightening discussions
regarding the subject.

\vskip 30pt
\centerline{\bf References}\parG

\item{[1]} A. Sommerfeld, {\sl Vorlesungen \"uber Theoretiche
Physik}, 2nd edition, DWB, Wiesbaden (1947) --- Chapter III.\par

\item{[2]} R.P. Feynman, R.B. Leighton and M. Sands, {\sl The Feynman
Lectures on Physics}, vol. I (exercises), Addison-Wesley, Reading,
Massachusetts (1964) --- exercise 4-12.\par

\item{[3]} E.A. Hylleraas, {\sl Mathematical and Theoretical
Physics}, vol. I, Wiley-Interscience, New York (1970) --- part II,
Chapter 27.\par

\item{[4]} P.A. Tipler, {\sl Physics}, vol. I, Worth Publisher, New
York (1976) --- Chapter 13, exercise 26.\par

\item{[5]} P.H. Lucie, {\sl F\'\i sica B\'asica}, vol. 2, Campus, Rio
de Janeiro (1980) --- Chapter 7, exercises 1R, 3R, 4R, 3, conceptual
questions 2, 3, 5, problems 1--7, 11 and 14.\par

\item{[6]} R.M. Eisberg, {\sl Physics: Foundations and Applications},
McGraw-Hill, New York (1981) --- Chapter 16, example 9 and problems
42--45.\par

\item{[7]} H.M. Nussenzveig, {\sl F\'\i sica B\'asica}, vol. 2, Edgar
Bl\"ucher, S\~ao Paulo (1983) --- Chapter 2, exercises 1--3, 6 and
7.\par

\item{[8]} J.C. Sartorelli, W.M. Gon\c calves and R.D. Pinto, {\it
Phys. Rev.} {\bf E49}(5), 3963-3975 (1994); X.D. Shi, M.P. Brenner
and S.R. Nagel, {\it Science} {\bf 265}, 219-222 (1994); T.J.P.
Penna, P.M.C. de Oliveira, J.C. Sartorelli, W.D. Gon\c calves and
R.D. Pinto, {\it Phys. Rev.} {\bf E52}(3), R2168-R2171 (1995); R.D.
Pinto, W.M. Gon\c calves, J.C. Sartorelli and M.J. de Oliveira, {\it
Phys. Rev.} {\bf E52}(6), 6896-6899 (1995); M.S.F. da Rocha, J.C.
Sartorelli, W.M. Gon\c calves and R.D. Pinto, {\it Phys. Rev.} {\bf
E54}(3), 2378-2383 (1996); J.G. Marques da Silva, J.C. Sartorelli,
W.M. Gon\c calves and R.D. Pinto, {\it Phys. Lett.} {\bf 226A},
269-274 (1997); W.M. Gon\c calves, R.D. Pinto, J.C. Sartorelli and
M.J. de Oliveira, {\it Physica} {\bf A257}, 385-389 (1998); R.D.
Pinto, W.M. Gon\c calves, J.C. Sartorelli, I.L. Caldas and M.S.
Baptista, {\it Phys. Rev.} {\bf E58}(3), 4009-4011 (1998); S.R.
Nagel, {\it Am. J. Phys.} {\bf 67}(1), 17-25 (1999); and references
therein.\par

\item{[9]} P.M.C. de Oliveira and T.J.P. Penna, {\it J. Stat. Phys.}
{\bf 73}(3/4), 789-798 (1993); P.M.C. de Oliveira and T.J.P. Penna,
{\it Int. J. Mod. Phys.} {\bf C5}(6), 997-1006 (1994); A.R. Lima,
T.J.P. Penna and P.M.C. de Oliveira, {\it Int. J. Mod. Phys.} {\bf
C8}(8), 1073-1080 (1997).\par

\item{[10]} S. Moss de Oliveira, P.M.C. de Oliveira and D. Stauffer,
{\sl Evolution, Money, War and Computers: Non-Traditional
Applications of Computational Statistical Physics}, Teubner,
Stuttgart-Leipzig (1999) --- Chapter 3.\par

\vskip 30pt
\centerline{Figure Captions}\parG

\item{Fig. 1} Experimental apparatus used in our test of Torricelli's
Law. Successive values of the height $H$ were read from marks
previously glued on the bottle surface, $0.5$cm apart from each
other. The successive speeds of the water squirted at the pin-hole
were indirectly measured through the ranges $X$ read at the
horizontal rule. Successive times $T$ were also measured during the
flow.\par

\item{Fig. 2} Two-bottles experiment, where the water squirted from
the first is caught by the second through a funnel. Successive
heights $H'$ were measured as a function of the time.\par

\item{Fig. 3} Squared range $X^2$ {\it versus} height $H$ for the
one-bottle experiment shown in figure 1. Our estimated error bars are
represented by the symbol's size. The continuous line is the best
linear fit for the experimental data. The inner diameter of the
bottle is 9.5cm, and that of the pin-hole tube is 2.0mm. The initial
height is $H_0 = 20.0$cm, according to which we measured the initial
horizontal range $X_0 = (28.0 \pm 0.3)\,$cm.\par

\item{Fig. 4} The same as in figure 3, now for a different bottle
with a different pin-hole. The inner diameter of the bottle is 5.1cm,
while that of the pin-hole tube is 3.5mm. The initial height is $H_0
= 42.0$cm, and the corresponding initial horizontal range is $X_0 =
(44.0 \pm 0.3)\,$cm.\par

\item{Fig. 5} Experimental speed {\it versus} height. Different
symbols correspond to the same experiment performed with different
bottles with different pin-holes, according to figures 3 (circles)
and 4 (crosses). By adopting the dimensionless, reduced variables $v$
and $h$, defined by equations (4) and (5), our results become
independent of the particular geometrical dimensions of the bottle
and the pin-hole: both data sets could be superimposed into the same,
universal, bottle-independent curve. The continuous line corresponds
to the theoretical Torricelli's law, equation (6).\par

\item{Fig. 6} Same experimental data of figure 5. Now, the continuous
line is a linear fit for the circles, excluding the last points of
the flow where instabilities and oscillations appear. The slopes
measured for each different bottle must be compared with the exponent
$1/2$ of the square root in theoretical equation (6).\par

\item{Fig. 7} Time evolution of the speed $v$. The continuous line is
the best fit excluding heights $H < 5$cm, below which an intermittent
flow regime starts to be observed.\par

\item{Fig. 8} Time evolution of the dimensionless speed and height,
as functions of the dimensionless time, equation (10). Experimental
and theoretical values agree very well with each other, except for
the last points where the flow is no longer stationary.\par

\item{Fig. 9} Experimental time evolution of the dimensionless speed
$u$ of the lowest bottle, as a function of the dimensionless time, in
the two-bottles experiment (figure 2). The continuous line comes from
the theory.\par

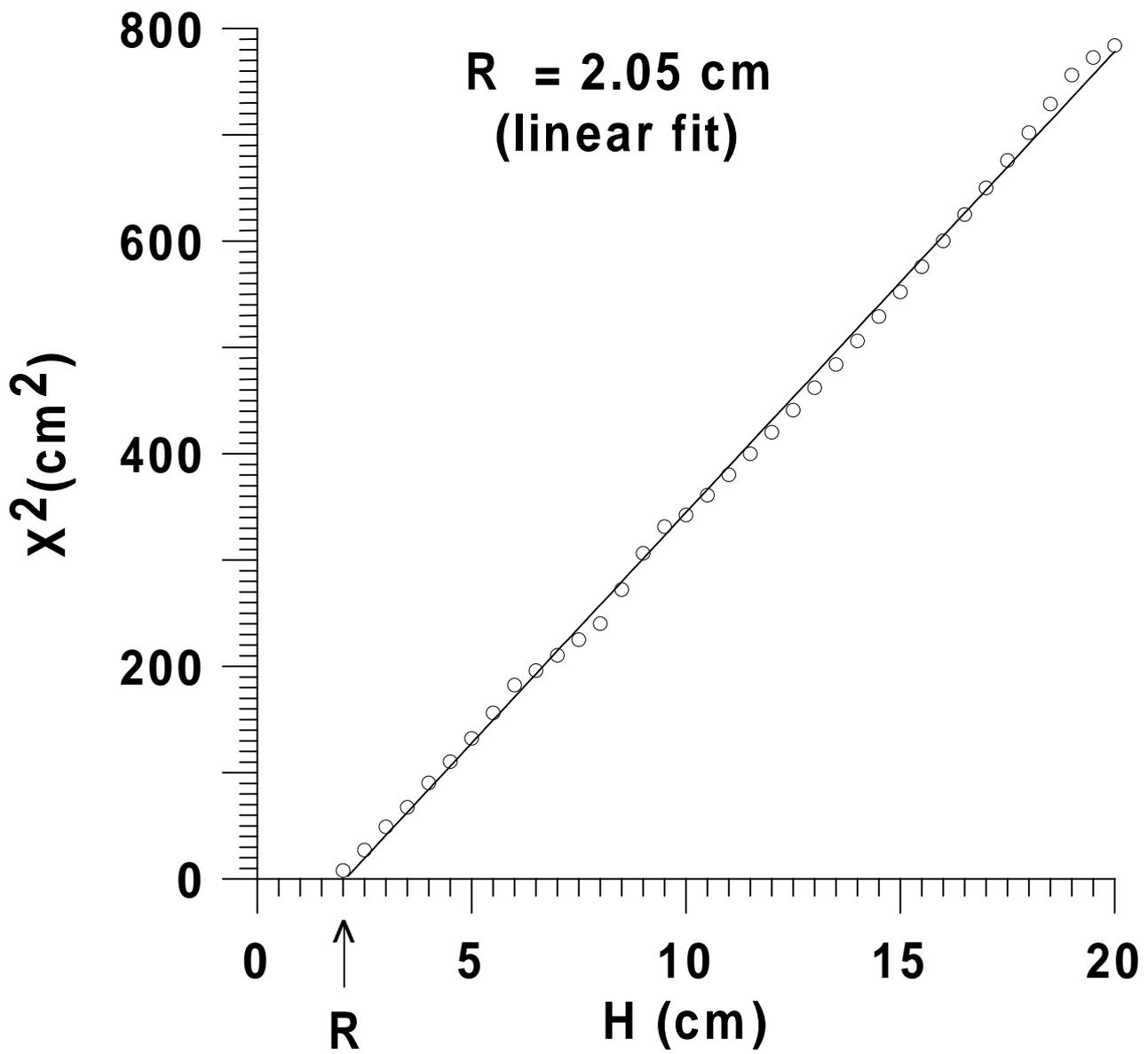
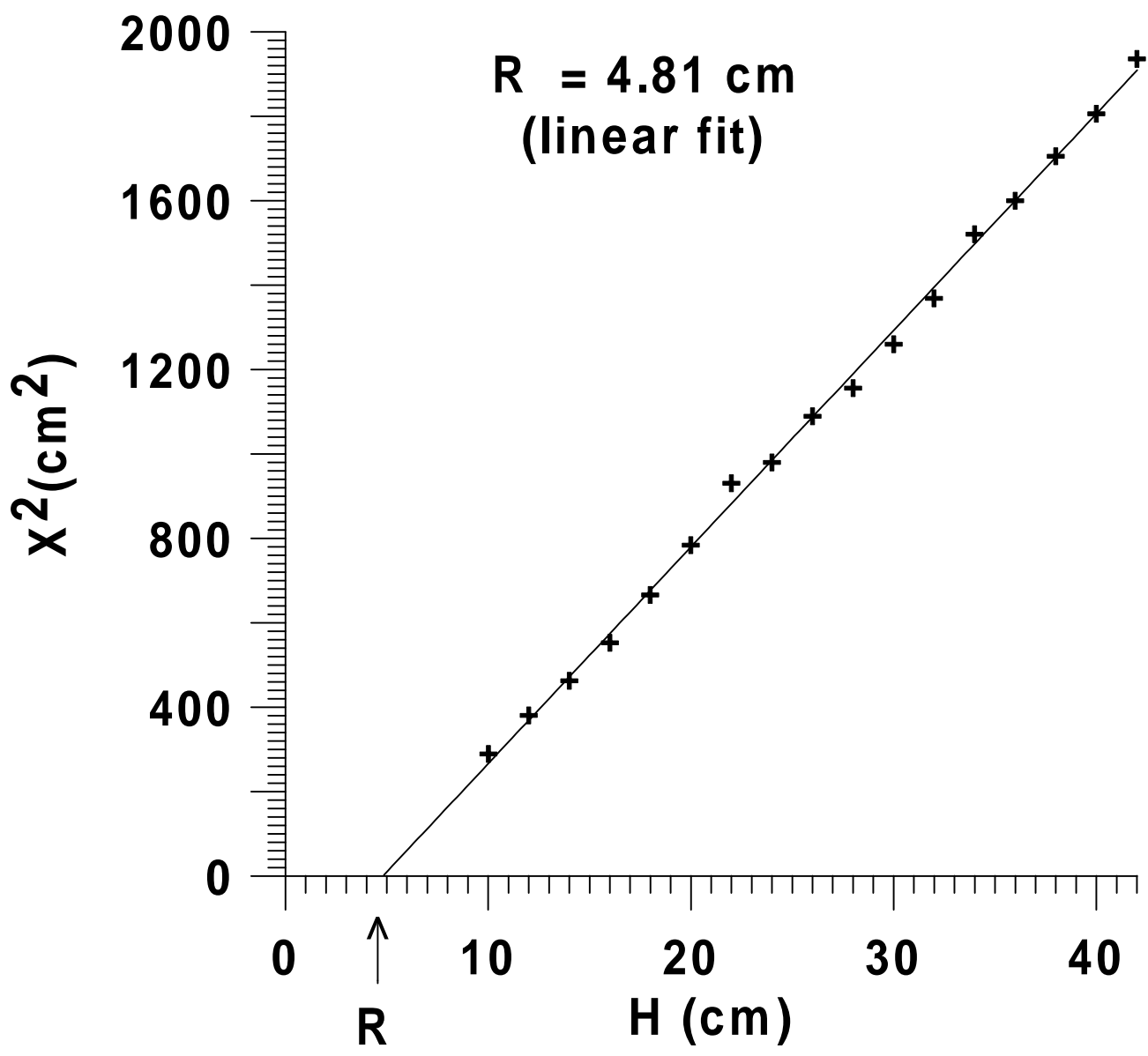
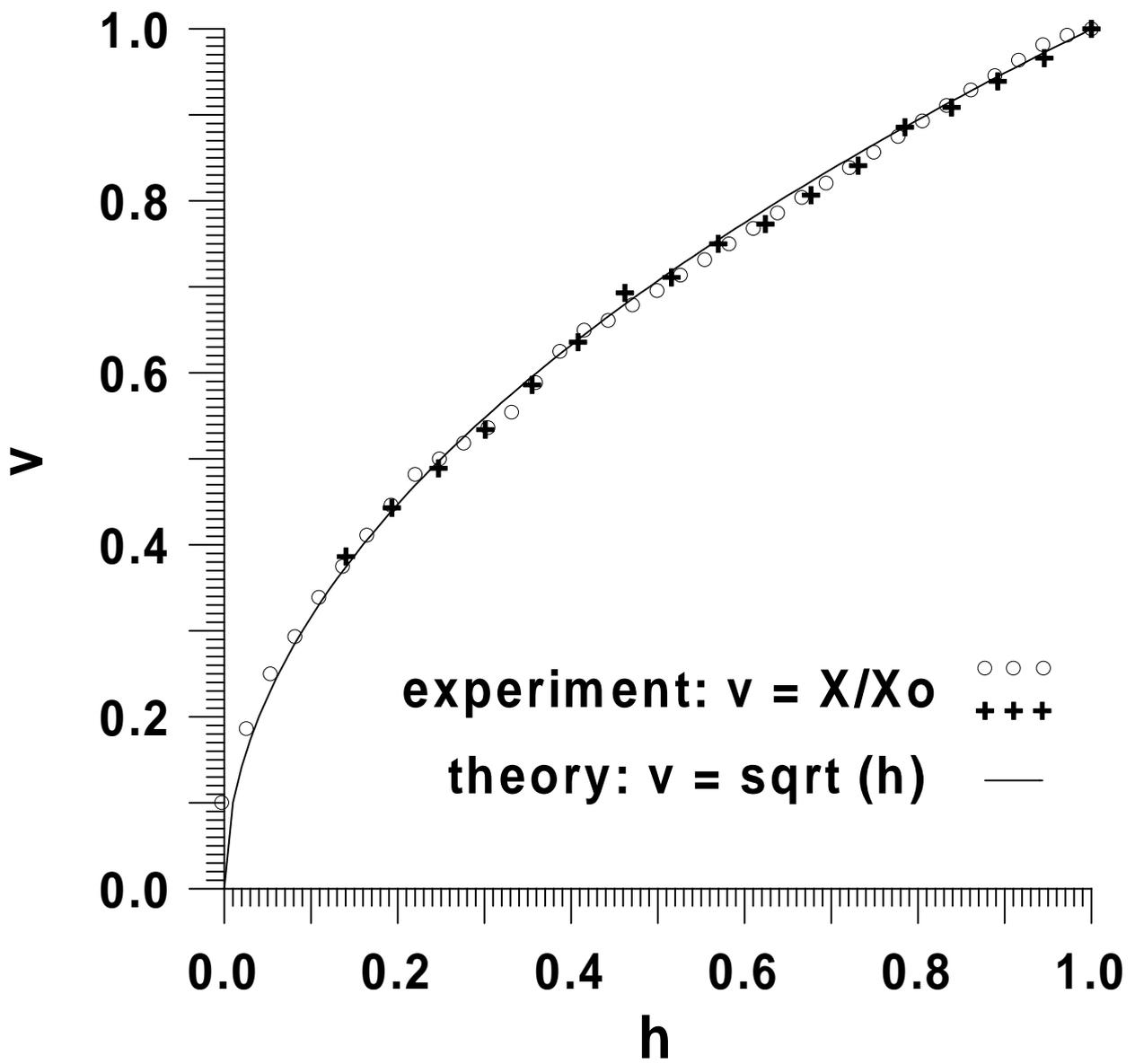
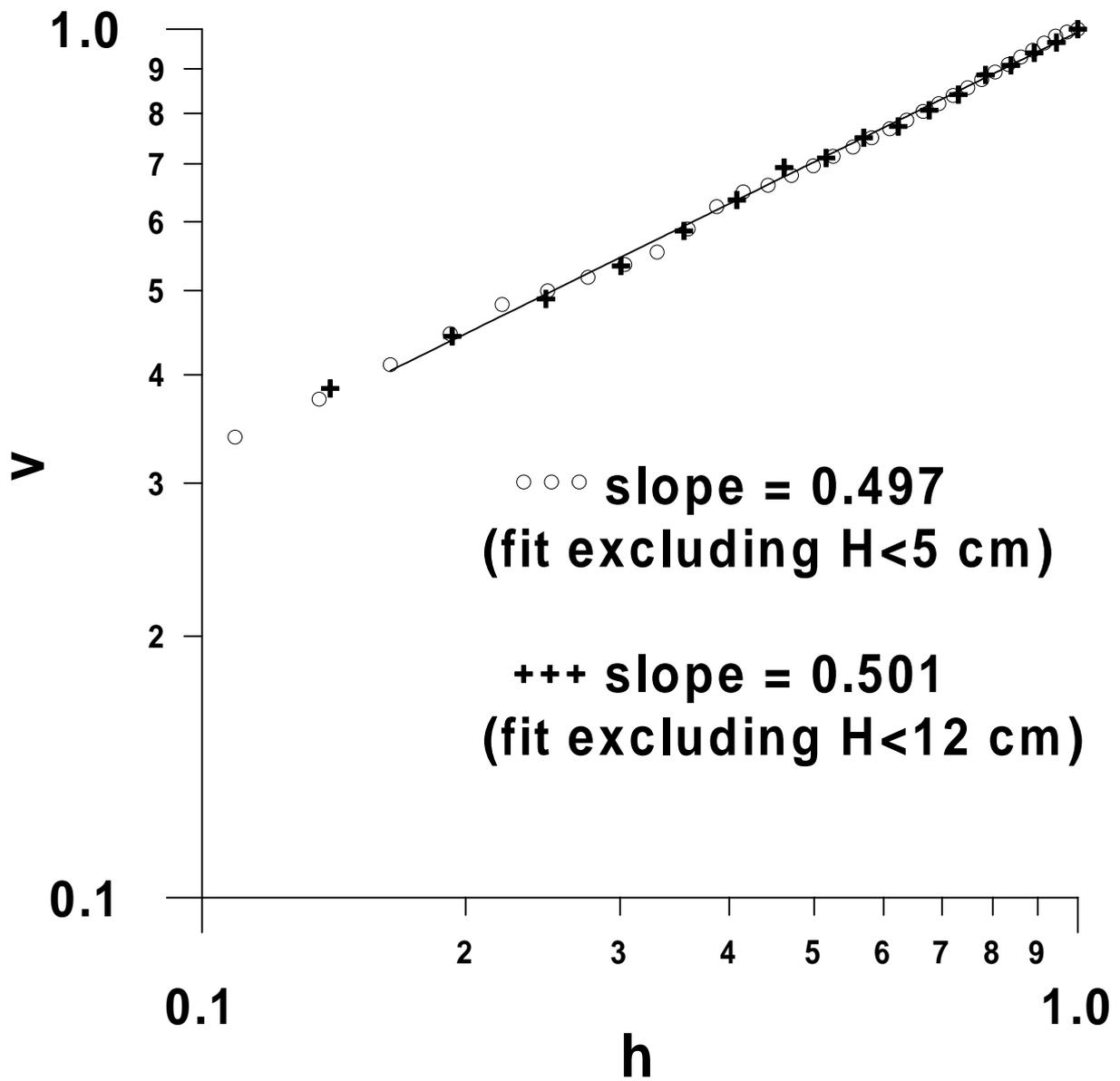
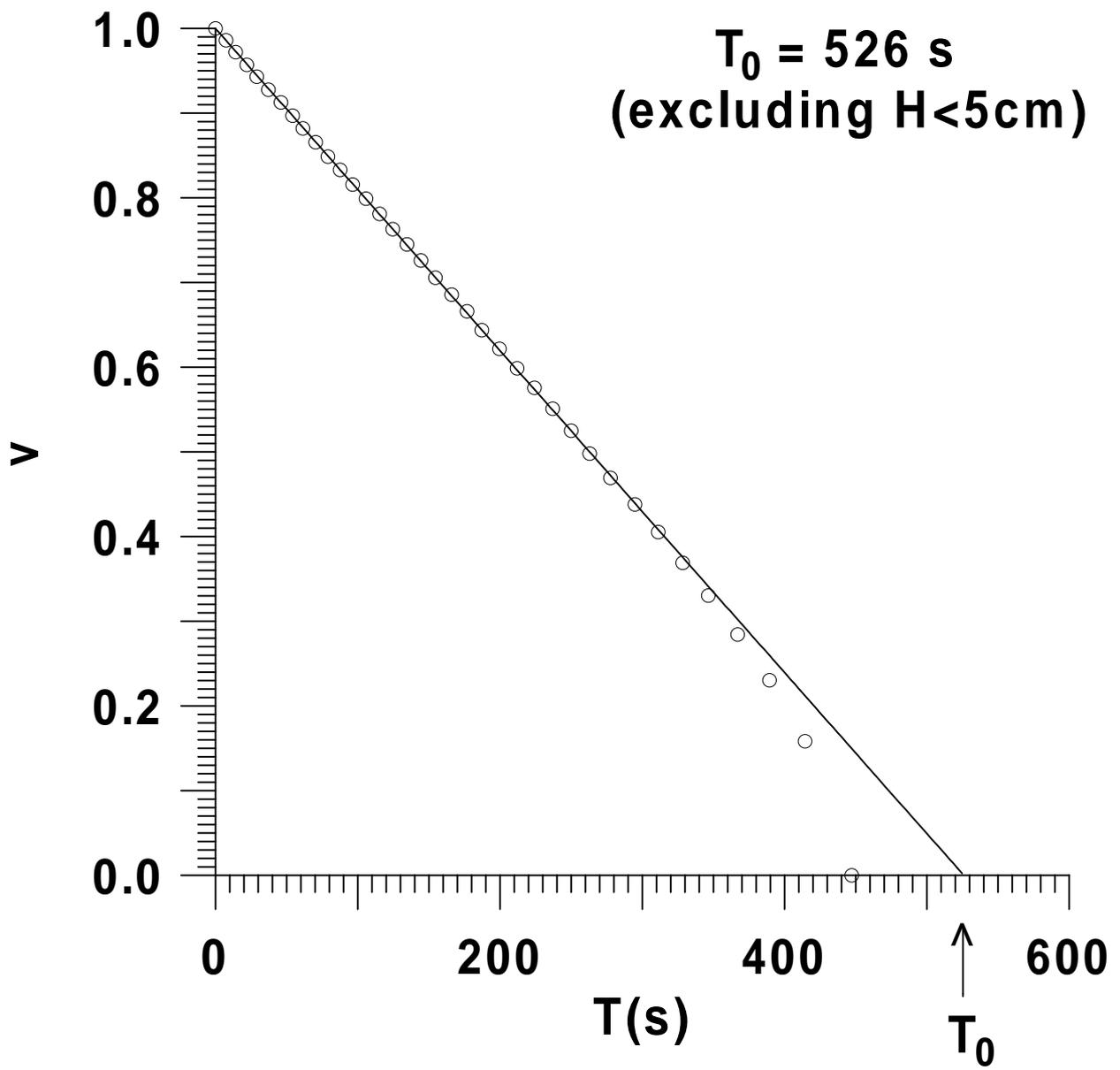
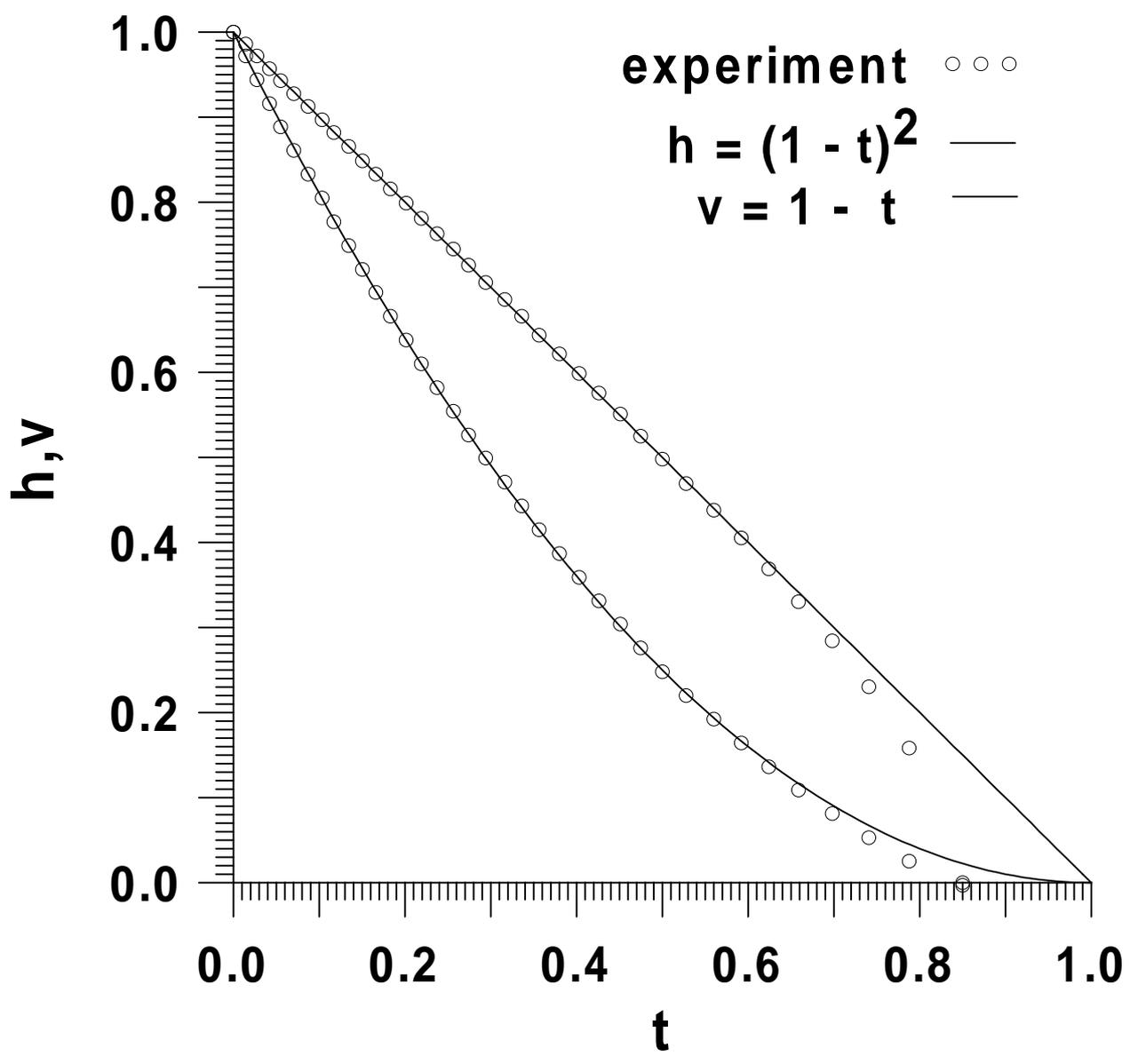
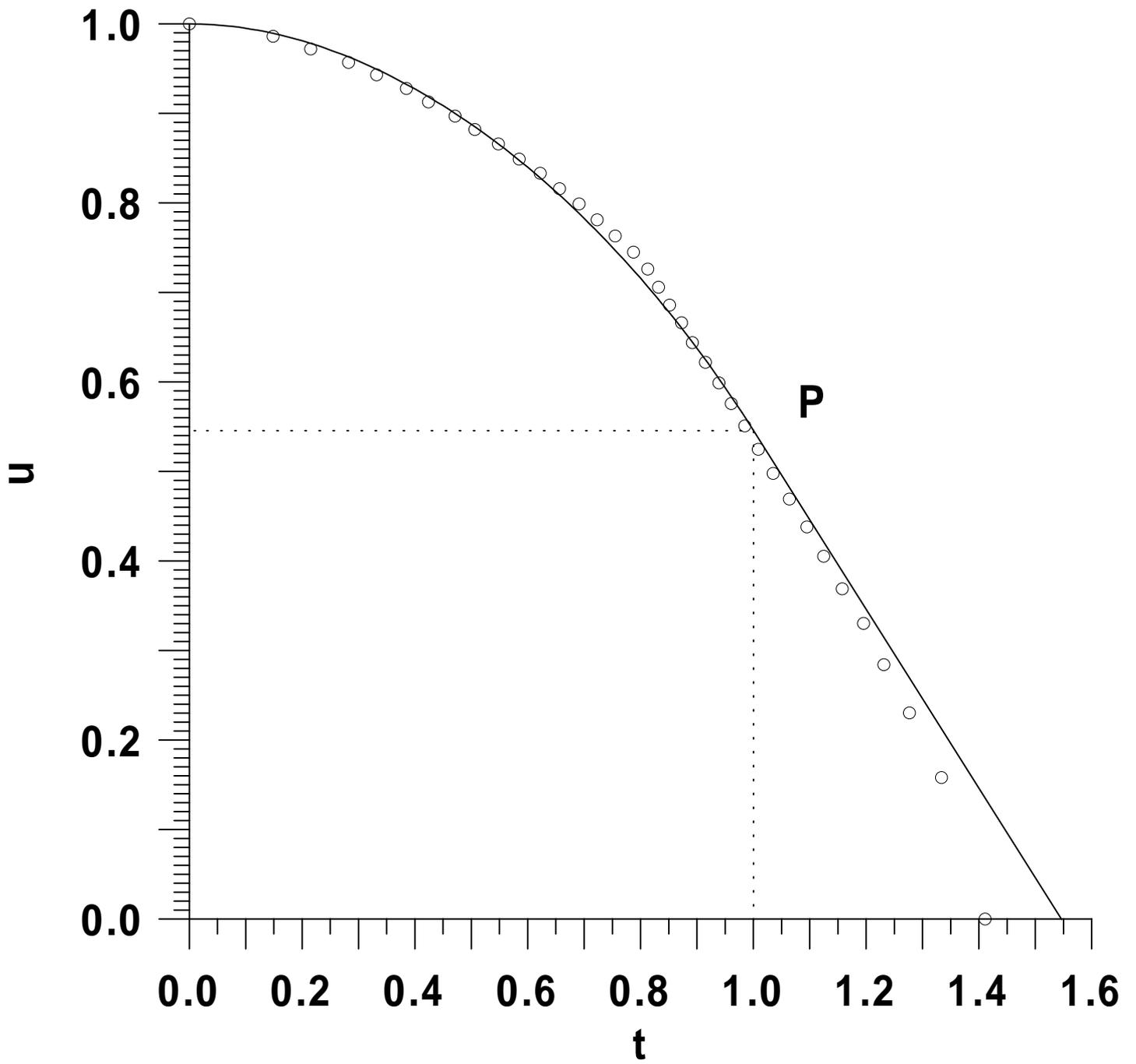
\bye